\documentclass[letterpaper]{article} 
\usepackage{aaai24}  
\usepackage{times}
\pdfoutput=1
\usepackage{helvet}  
\usepackage{courier}  
\usepackage[hyphens]{url}  
\usepackage{graphicx} 
\usepackage{natbib}  
\usepackage{caption} 
\DeclareCaptionStyle{ruled}{labelfont=normalfont,labelsep=colon,strut=off} 
\usepackage{algorithm}
\usepackage{algorithmic}
\usepackage{amsmath}
\usepackage{amsfonts}
\usepackage{newfloat}
\usepackage{listings}
\floatstyle{ruled}
\newfloat{listing}{tb}{lst}{}
\floatname{listing}{Listing}
\usepackage{graphicx, color, xcolor, colortbl}

\begin{document}
%
\title{Promoting Fair Vaccination Strategies Through Influence Maximization: A Case Study on COVID-19 Spread}
\author {
    Nicola Neophytou\textsuperscript{\rm 1,2},
    Afaf Ta\"ik\textsuperscript{\rm 1,2},
    Golnoosh Farnadi\textsuperscript{\rm 1,2,3}
}
\affiliations {
    \textsuperscript{\rm 1}Mila, Quebec AI Institute, Quebec, Canada, \\
    \textsuperscript{\rm 2}Université de Montréal, Quebec, Canada, \\
    \textsuperscript{\rm 3}McGill University, Quebec, Canada\\
    nicola.neophytou@mila.quebec, afaf.taik@mila.quebec, farnadig@mila.quebec
}

\setlength{\floatsep}{0mm}
\setlength{\textfloatsep}{0mm}
\setlength{\dblfloatsep}{0mm}
\setlength{\dbltextfloatsep}{0mm}
\setlength{\abovedisplayskip}{0mm}
\setlength{\abovedisplayshortskip}{0mm}

\maketitle

\begin{abstract}
The aftermath of the Covid-19 pandemic saw more severe outcomes for racial minority groups and economically-deprived communities. Such disparities can be explained by several factors, including unequal access to healthcare, as well as the inability of low income groups to reduce their mobility due to work or social obligations. Moreover, senior citizens were found to be more susceptible to severe symptoms, largely due to age-related health reasons. Adapting vaccine distribution strategies to consider a range of demographics is therefore essential to address these disparities. In this study, we propose a novel approach that utilizes influence maximization (IM) on mobility networks to develop vaccination strategies which incorporate demographic fairness. By considering factors such as race, social status, age, and associated risk factors, we aim to optimize vaccine distribution to achieve various fairness definitions for one or more protected attributes at a time. Through extensive experiments conducted on Covid-19 spread in three major metropolitan areas across the United States, we demonstrate the effectiveness of our proposed approach in reducing disease transmission and promoting fairness in vaccination distribution.
\end{abstract}

\maketitle

\section{Introduction}
\begin{figure}
    \centering
    \includegraphics[width=7.8cm]{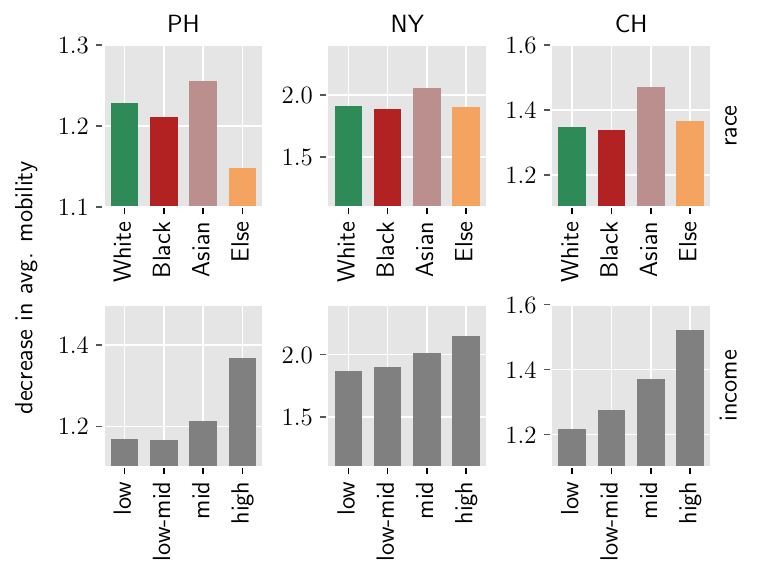}
    \caption{The \textit{reduction} in mobility from before lockdown to during lockdown per racial groups and income groups in Philadelphia, New York and Chicago metropolitan areas. For all three areas, lower income groups and racial minorities belonging to lower income groups (see Fig. \ref{fig:raceincome}) were less able to reduce their mobility as quickly when transitioning to lockdown.}
    \label{fig:mobility-reduction}
\end{figure}

The fallout of Covid-19 revealed the stark inequalities in access to healthcare between social groups in diverse and urban areas \cite{Ndugga, Price-Haywood2020, Millett2020, Joseph2020, Azar2020, Hsu2020, Gayam2021}. Studies confirmed that economically-deprived communities and racial minorities experienced higher rates of infection, hospitalization and mortality as a result of Covid-19 \cite{Kirby2020evidence, Tai2021disproportionate, Alcendor2020racial}.
The reasons for this disparity form a long chain of events, with unequal access to healthcare between socioeconomic groups, and therefore racial groups, at the root of it. Furthermore, studies on mobility networks in the US also revealed how minority communities were less able to reduce their mobility as quickly during the pandemic, and as a result suffered higher rates of infection \cite{Chang2020}. This can be largely attributed to underprivileged groups assuming the roles of frontline and critical infrastructure work, and also living and working in more crowded circumstances. Inequalities in access to the internet and ease of travelling to test and vaccination sites are also factors contributing to this discrepancy \cite{Yee2022}. These data demonstrate inequities in receiving resources throughout the pandemic, which also extends to vaccination \cite{Perry2021, Kates, KFF, Bayati2022inequality}. In Figure \ref{fig:mobility-reduction}, we demonstrate how racial minorities and lower income communities in three US metropolitan areas were less able to reduce their mobility as quickly when the lockdown was introduced.

This disparity motivates the need for a fair vaccination strategy that differs from the current technique.
In this work, we investigate a collection of alternative vaccination strategies that consider both mobility and fairness. We leverage an approach called influence maximization (IM), a network science technique designed to detect the most influential members of social networks, typically used in applications such as viral marketing campaigns \cite{Chen2010}. We adapt this principle to instead detect the neighborhoods or communities which exhibit the largest influence on a mobility network in terms of disease propagation. Such communities are likely to include essential workers who are less able to reduce their mobility during lockdowns \cite{rasnavca2020essential,NGUYEN2020e475}.

We adapt our IM approach to achieve fairness in vaccine allocation for racial groups as well as groups of different social statuses. Moreover, older individuals may be less mobile but more at risk of severe outcomes when exposed to the disease. It is essential not to overlook this trade-off; we therefore also design a strategy designed to protect communities based on higher risk and vulnerability.
To summarize, our contributions are as follows:
\begin{enumerate}
\item A novel community-level influence maximization approach for identifying impactful neighborhoods, aiding targeted vaccination against disease transmission.
\item Extension of influence maximization to mitigate infection disparities among racial and income-level communities.
\item Introduction of a competitive method, merging influence maximization with prioritizing older communities to reduce overall infections.
\item Empirical validation on mobility networks from three major US metropolitan areas, utilizing real aggregated visit data from census block groups (CBGs) to points of interest (POIs) during the first five weeks  of Covid-19 pandemic.
\end{enumerate}

\section{Related Work}
Fairness in AI is an expanding area of research which has seen traction since the exposure of biases in several significant technologies \cite{Mehrabi2021survey, gendershades, compas}. This extends to the field of influence maximization, where various works impose fairness constraints on the optimization problem. For instance, Farnadi et al. provide a framework for applying a variety of fairness definitions to IM tasks \cite{Farnadi2020}. Other works have included developing adversarial graph embeddings to achieve fair IM in social networks \cite{Khajehnejad2020adversarial}, and balancing majority and minority groups in IM when networks and diffusion processes demonstrate homophily \cite{Anwar2021}. Ali et al. ensure fairness in exposure under a time-critical perspective, for example posting a job advertisement which should be reached by equal sub-populations before the deadline to apply for the position \cite{Ali2021}. This does not however consider the possibility of changing network structure over time, as is the case with mobility or contact networks used in epidemiology, and is therefore not extendable to the problem of vaccine distribution. 

Meister et al. characterize communities by their social activity and vulnerability due to age in a comparative game for optimizing vaccination, however, they do not use real contact or mobility data to test their approach \cite{Meister2021fair}.
Similarly, works which do consider demographic fairness in vaccination do not necessarily consider mobility to improve performance \cite{Kipnis2021evaluating, nationalacademics2020, Kadelka2022ethnic}. Anahideh et al. propose a vaccine allocation solution which tackles the apparent trade-off between equal distribution amongst regions and demographic fairness \cite{Anahideh2022fair}. Their approach defines effective distributions as those that cover many geographical areas; however, we argue that this does not necessarily reduce infections most effectively as it treats each area as equally influential in disease transmission. Commonly, works that do consider social contact often do not consider fairness implications \cite{Chen2021prioritizing}. Much research considers how to optimize the age-based vaccination approach, but often other sensitive attributes are not considered \cite{Wu2022how, Sheldrick2022plausibility, Kirwin2021netbenefit, Ferranna2021Covid, Gozzi2022estimating, Sheldrick2022plausibility, Jentsch2021, Buckner2021}.

Similar to our approach, Minutoli et al. \cite{Minutoli2020preempt} lay the theoretical foundations for using IM to reduce disease transmission via vaccination, though they do not consider fairness. They use a simplistic propagation model which is not specific to a particular disease, nor does it consider important factors in disease transmission such as the number of people in a confined space, or the duration spent there, as we do. Additionally, their implementation is for contact networks containing interactions between individuals, rather than communities. We argue that privacy regulations on mobile tracking data will limit the ability to reconstruct such specific interactions, not to mention also matching demographic information to individuals in these interactions. Using aggregate visits as in our work is therefore more realistic in terms of the data available, and allows us to extract demographic information on the community level which we use for our fairness approaches. Rather than using large contact networks, condensed mobility networks of aggregate visits also drastically reduces the network size which is beneficial for computation. \\
Mehrab et al. use the same mobility data as in our contribution to guide vaccine distribution to groups with lower up-take rates \cite{Mehrab2022datadriven}. However, they use visit counts to public places to determine the best candidates, whereas we determine influence by simulating disease spread on top of the mobility network to predict infections. Further, their contribution is to identify groups with low vaccine uptake at a later stage of vaccine deployment, whereas we offer a solution for the first stages of the allocation, to ensure fairness from the beginning. To our knowledge, our work is the first approach to use \textit{community-level} influence maximization to propose vaccination strategies which consider demographic fairness.

\section{Preliminaries}
\subsection{Influence Maximization}
IM is a network science technique used to identify the most influential nodes in a graph, with respect to their ability to propagate a certain transmissible quality, such as information or disease. IM is particularly popular in viral marketing problems on social media platforms, in which an entity wants to share advertisements with only a few individuals online, but hopes to optimize this choice by selecting those who are more likely to share the information with the rest of the network. 

While some techniques rely on heuristics like node centrality or betweenness to measure a node's importance in a graph \cite{Newman2005betweenness, Borgatti2005centrality}, IM assumes access to a function which models the propagation of the given substance across the network. The algorithm uses this propagation function to generate its selection for the set of most influential nodes, often referred to as the seed set. In their seminal work, Kempe et al. \cite{Kempe2015} proposed a greedy algorithm which achieves $(1-\frac{1}{e})$ optimality guarantees on the seed set, so long as the objective function - which is the given propagation function - is both submodular and monotonic.
However, the biggest drawback of the greedy strategy is its inefficient runtime. The commonly used propagation functions, such as the Linear Threshold \cite{LTmodel} and Independent Cascade \cite{ICmodel} models, are stochastic in nature. The greedy strategy, therefore, relies on running several Monte Carlo simulations of these stochastic propagation functions, which is costly. Much of the proceeding work following this focus on trying to improve runtime performance; CELF exploits the submodularity of the influence function to reduce the number of influence evaluations, achieving a runtime of up to 700 times faster than the greedy algorithm \cite{Leskovec2007}. In our experiments, we also use CELF to reduce the number of evaluations.

\subsection{Mobility networks}
We draw on the work conducted by Chang et al. to construct our mobility networks and simulate Covid-19 propagation on them \cite{Chang2020}. The mobility network of a metropolitan statistical area (MSA) contains $K$ nodes which represent CBGs, neighborhoods of a few hundred to a few thousand residents. Each CBG is denoted by $c_i$, with $i=1,..,K$.
The population of each CBG is known and is given by $n_{c_i}$ for CBG $c_i$. The total number of residents in the network is, therefore, $N = \sum_{i=1}^K n_{c_i}$. Each individual in the network can belong to one of $M$ social groups (e.g. racial groups) indexed by $j=1,..,M$. In our work, we experiment with $j$ representing racial groups and also groups based on their median household income, obtained from US Census data \cite{Census}. For racial groups, each CBG can contain any number of individuals belonging to each social group, up to its total population size. The fraction of residents in CBG $c_i$ who belong to racial group $j$ is known and given by $\alpha_{ij}$, where $0 \leq \alpha_{ij} \leq 1$, and therefore $\sum_{j=1}^M \alpha_{ij} = 1 \forall i$. The total number of individuals in the network belonging to racial group $j$ is therefore obtained by $N_j = \sum_{i=1}^K \alpha_{ij}n_{c_i}$. For social status groups, we split the CBGs into $M$ groups defined by the median income of the households in that CBG. The total number of individuals belonging to a group is, therefore, $N_j = \sum_{i=1}^K n_{c_i}$ if $c_i$ belongs to income group $j$.

\subsection{Covid-19 propagation model}
We use the Covid-19 simulation proposed by Chang et al. \cite{Chang2020}, which models the propagation of the disease on the network of CBGs and POIs. The constructed hourly visits from the network describe how many individuals travel from CBGs to POIs per hour. POIs here represent public places such as restaurants, gyms and religious centers, where interactions with members of other communities can happen, and the disease can spread accordingly.
Individuals can also transmit the disease amongst members of their home CBG. The number of new infections per CBG is therefore a summation of two terms drawn from different distributions; one Poisson distribution for new exposures from POIs, and one Binomial distribution for new exposures from their home CBG and other places which may not be accounted for in the POIs, such as public transportation. The previous work fixed the parameters of the model by comparing it to the counts of real Covid-19 cases, which we also reproduce in our work. Our vaccination approach can technically be used to combat any disease, simply by swapping the influence function $\sigma$ for a model of said disease. This would mean changing the parameters of the model when calibrating it to real case counts. Further details of the model and its calibration can be found in the Appendix. \\
At all time steps in the disease simulation, we have access to the number of susceptible, infected, exposed, and recovered or removed individuals residing in each CBG. We maintain a vector of size $K$ for each of these states, denoted respectively by $S$, $E$, $I$, and $R$. Their elements are indexed by $i$ and contain the fraction of individuals in CBG $c_i$ belonging to that state.
For example, element $I_i^t$ contains the fraction of individuals in CBG $c_i$ who are in the infected state at time $t$. In this work, we are only interested in the final rates of exposed, infected or recovered/removed (EIR) individuals in the final time step of the model, $T$, and readers can assume we use the final iteration at $T$ of these vectors from here onwards. For example, the sum of all exposed-or-worse individuals in the network at time $T$ is given by $N_{EIR} = \sum_{i=1}^K (E^T_i + I^T_i + R^T_i)n_{c_i}$. \\
Individuals in the network therefore always belong to one of $M$ social groups, as well as one of the four SEIR states. Since we only have access to these statistics on the community level, we approximate the number of people in the network belonging to racial group $j$ and exposed-or-worse as $N_{EIR_j} = \sum_{i=1}^K (E^T_i + I^T_i + R^T_i) \alpha_{ij}n_{c_i}$. For income group $j$, the equivalent is obtained by $N_{EIR_j} = \sum_{i=1}^K (E^T_i + I^T_i + R^T_i)n_{c_i}$ for CBGs belonging to income group $j$.

\section{Proposed Approach}
\label{sec:methodology}
In this section, we present three methods of targeted vaccination using IM. Firstly, we present our simple method for vaccinating with IM and no additional constraints. We then outline two methods for introducing fairness to IM, for both racial groups and income groups. Finally, we present a method for applying weights to communities corresponding to their relative risk, in order to use IM and still prioritize older communities that are more vulnerable to severe outcomes if infected.\\
In our approach, the treatment of the three sensitive attributes (also referred to as protected attributes) in the network - race, income level and age - are not the same. For the racial and income groups, we aim to achieve fairness according to their population size in the network. An individual should be no more at risk of infection due to their race or income than what is expected given the racial or income group's population size in the network. However, the same strategy should not be adopted for age, since infection can lead to more severe outcomes for older individuals, making them more high risk. Therefore, we strive to achieve fairness amongst race and income, but adopt a \textit{bias} with respect to age, in order to protect individuals at high-risk.

As addressed in the Preliminaries, the greedy approach in influence maximization provides provable guarantees on the optimality of the seed set, so long as the influence function is both monotonic and submodular. Previous findings have found issues proving these properties in a temporal SIR model \cite{Erkol2022submodularity}. Similar to this work, we argue that the greedy approach is still effective despite the propagation model exhibiting submodularity violations. Additionally, the greedy approach is more scalable for greater network sizes than solving the optimization problem exactly, particularly when fairness constraints are required \cite{Farnadi2020}. Further, we argue that the greedy approach provides an element of model interpretability, which is particularly important when justifying why one neighborhood should receive vaccines over another; in our case, we can clearly demonstrate with the greedy approach that a neighborhood with higher mobility and influence is more likely to be selected for vaccination. \\
Further, we argue that using community-level influence maximization, with aggregated data on CBGs rather than individuals, is a more realistic approach when we want to obtain demographic information for fairness purposes. Privacy concerns (rightly) limit access to fine-grained data on individual mobility and their sensitive attributes, but here we use Census data to match demographic data to CBGs.

\subsection{Vaccinating with Influence Maximization}
In this section, we outline how to select the most influential communities in the network in terms of disease spread using IM and CELF. The algorithm is provided in Algorithm \ref{alg:im}.
We begin with a budget $B$ corresponding to the number of vaccines available for allocation to the whole network. We can simulate the disease spread from a set $Z$ of CBGs, in order to quantify how influential those communities are. In the general case, $Z$ can contain any number of CBGs, but to quantify the influence of just one, only that CBG would be contained in the set $Z$, e.g. $Z = \{c_i\}$ for $i=1,..,K$. 
For all $K$ CBGs in the network, we then conduct simulations of the disease spreading from that neighborhood alone.
We maintain a list of lists $L = [[1, \sigma(\{c_1\})],...,[K, \sigma(\{c_K\})] ]$, where for CBG $i$, $\sigma(\{c_i\})$ represents the $N_{EIR}$ count as a result of simulating disease spread starting only from $c_i$. $L$, therefore, contains the list of pairs of the candidate CBGs indices along with their corresponding influence, as a count of how many people resulted in exposed-or-worse states.\\
We initialize the set of nodes to vaccinate, $Z$, as an empty set. Then, in each iteration, the CBGs with the greatest marginal gain are greedily added to $Z$. The marginal gain is the difference in influence between the current set of selected CBGs ($spread$), and the influence of the current selected CBGs \textit{plus a potential candidate CBG}. Note that the gain is also normalized by the population of the CBG, $n_{c_i}$. We implement this normalized version of IM since we want to select CBGs that are the most influential per their population, and CBGs with a higher population use more of the vaccination budget than CBGs with a lower population. We keep track of how much budget is used so far with $B'$, which gets updated with the population sizes of CBGs when they are added to $Z$. 

\begin{algorithm}[tb] 
\caption{Selecting CBGs to vaccinate using IM and CELF}
\label{alg:im}
\textbf{Input}: budget $B$, number of CBGs $K$, disease model $\sigma$\\
\textbf{Output}: CBGs to vaccinate $V$
\begin{algorithmic}[1]
\STATE $B' \gets 0, spread \gets 0, L \gets [], Z \gets []$
\FOR{$i=1, \dots, K$}
    \STATE $L$.append($[i, (\sigma(\{c_i\}) - spread)/n_{c_i}]$)
\ENDFOR
\STATE sort $L$ by gain, descending
\STATE $Z$.append($L_{0,0}$)
\COMMENT{add cbg with best gain to Z}
\STATE $spread \gets L_{0,1}*n_{Z_{-1}}$
\STATE $B' \gets n_{Z_{-1}}$ 
\COMMENT{update budget used}
\WHILE{there are possible candidates in $L$}
    \STATE $matched \gets False$
    \WHILE{not matched}
        \STATE $best \gets L_{0,0}$
        \STATE $L_{0,1} \gets (\sigma(Z \cup \{best\}) - spread)/n_{best}$
        \STATE sort $L$ by gain, descending
        \STATE $matched \gets L_{0,0}$ is $best$
    \ENDWHILE
    \STATE $spread \gets spread + L_{0,1}*n_{best}$
    \STATE $Z$.append($L_{0,0}$)
    \STATE $B' \gets B' + n_{Z_{-1}}$
    \COMMENT{update budget used}
    \STATE $L \gets L[1:]$
    \COMMENT{remove best from the candidate list}
    \STATE keep only candidates in $L$ which cannot exceed B
\ENDWHILE
\STATE \textbf{return} $Z$
\end{algorithmic}
\end{algorithm}

In lines 9-15, we perform a check to test whether the highest-influence candidate after the previous iteration is still the highest-influence candidate in the current iteration. If this is true, we omit the requirement to re-calculate the influence of the other candidates. This exploits the submodularity of the influence function, since the marginal gain of adding CBG $c_i$ to a smaller set $Z$ can only decrease. This is the contribution made by CELF \cite{Leskovec2007}, which we use to improve run time. The algorithm then continues to add candidate CBGs to $Z$ so long as their addition does not exceed the budget $B$. The final output of the model is therefore a set of CBGs to be vaccinated, which we call $V$. Our subsequent variations of this contribution in the next sections adapt this method to apply demographic fairness.

\subsection{IM with equal treatment} \label{method:eqtreatment}
Equal treatment is an existing fairness notion in the domain of fair IM, which aims to achieve fair representations of social groups in the final set of selected nodes $V$. This is equivalent to achieving the same demographic distribution in the set of communities to be selected for vaccination as in the whole network.

We model this task as a multiple knapsack problem \cite{multipleknapsack}, whereby each social group $j=1,..,M$ is allocated a number of vaccines based on the fraction of their population in the network. Each social group $j$, therefore, receives its own budget $B_j$, corresponding to the number of vaccines to be allocated to the group, given by 
\begin{equation}
    B_j = \frac{N_j}{N}B
\end{equation}
where $N_j$ is the number of individuals in the network belonging to group $j$ and $N$ is the total network population. To implement this, when a CBG is selected for vaccination, we update the budget used by each of the $M$ social groups, between lines 19 and 20 of Algorithm \ref{alg:im}. Additionally, after line 21, we perform another check to ensure the remaining candidate list contains only CBGs whose addition would not violate any of the social group budgets $B_j$. We use these definitions to outline two strategies: equal treatment by racial groups and equal treatment by median household income.
\subsubsection{\textbf{Equal treatment by racial groups (IM-R)}}
When performing equal treatment for racial groups, the budget of racial group $j$ is updated when CBG $c_i$ is selected for vaccination via
\begin{equation}
    B_j' = B_j' + n_{c_i}\alpha_{ij}
\end{equation}
This update is performed for all $M$ racial groups when any additional CBG is selected. 

We acknowledge that our fairness framework takes a Western-centric perspective, particularly with respect to race, and is relevant mostly to countries containing urban areas with high diversity, as is more typical of the global West. The most effective strategy for each country, however, will not be the same \cite{Eyal2022coronavirus}. We therefore propose IM-I to counteract this bias; many non-Western countries may not have the same extent of racial diversity, but will still experience income disparities in their urban areas. Using income as a sensitive attribute therefore still provides a fair IM method which is relevant for areas with low racial diversity. Further, Figure \ref{fig:raceincome} identifies that in our selected MSAs, the White population tends to dominate higher income CBGs, while historically marginalized groups of Black or African-American are more prevalent in lower income groups, as is typical of high-diversity urban areas in the West. Therefore, fairness by income level may also achieve fairness by racial groups for our selected MSAs.
\subsubsection{\textbf{Equal treatment by median income (IM-I)}}
We perform a similar equal treatment scenario, this time with social groups defined by income. We use labels of the median household income of each CBG. The distribution of the CBG median income is split into four quartiles. We then bucket the CBGs into one of $M=4$ groups according to which quartile its median income falls into. The budget is split in the same way as with race, using $B_j = \frac{N_j}{N}B$ for each income group $j$, where $N_j$ is the total population of that income group. However this time, the budget updates are given by
  \begin{equation}
    Bj' =
    \begin{cases}
      B_j' + n_{c_i}, & \text{if $c_i$ belongs to group $j$} \\
      B_j', & \text{otherwise}
    \end{cases}
  \end{equation}

\subsection{IM with age-associated risk-weights (IM-A)} \label{method:riskweights}
While older individuals in the network are less mobile, they are more likely to experience severe consequences if they contract the disease, including hospitalisation and death. As a result, it is important to consider this tradeoff when using IM. To do so, we incorporate this notion into our IM technique such that the significance of infecting a person from a CBG with a higher median age is greater. We implement this by weighting the CBGs with a ``risk-factor" according to their median age. The purpose is to not only select CBGs for vaccination that are highly influential in the network but in particular, find those CBGs that pose more of a risk of exposing older communities to the disease.
We achieve this by scaling the influence calculations $\sigma$, which is used as the selection criteria for vaccination (see Algorithm \ref{alg:im}). We construct a vector $\mu$ of size $K$ containing the associated risk-weights for each CBG, determined by its median age. We fill this vector with the death rates per age group from the CDC, which are the rates of death of older groups compared to 18-29 year olds in the US \cite{cdc-risks}. We report these values in the Appendix. The resulting influence, which before was just the sum of exposed-or-worse individuals in the network, is now instead a weighted sum where each CBG is weighted according to its risk-factor in vector $\mu$, given by
\begin{equation}\label{eq:risk}
    \sigma_A = \sum_{i=1}^{K} \mu_in_{c_i}(E_i + I_i + R_i).
\end{equation}
We use this metric as a proxy to account for infections as well as more severe cases and deaths. In Algorithm \ref{alg:im}, the influence function $\sigma$ is replaced by $\sigma_A$ in lines 3 and 13.

\begin{figure}
    \centering
    \includegraphics[width=8.2cm]{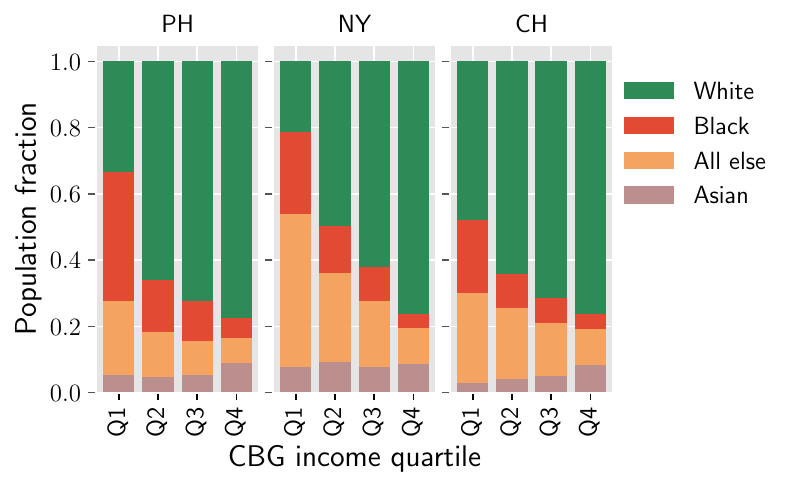}
    \caption{Racial distributions of CBGs grouped by their median income. Income groups are determined by quartiles of the median income distribution. Results are for three MSAs: Philadelphia, New York and Chicago.}
    \label{fig:raceincome}
\end{figure}

\subsection{Multiple protected attributes}
The methods proposed so far each strives for fairness using one protected attribute at a time. However, it is possible that achieving the desired fairness for one protected attribute is not beneficial for another. For example, older communities tend to be predominantly White in the US. Therefore, a vaccination strategy based solely on age, without consideration for fair distribution amongst racial groups, leads to an unfair allocation. This would also lead to unfair allocation amongst income groups as White populations tend to dominate higher income CBGs, as shown in Figure \ref{fig:raceincome}. To address this issue, we propose the following combinations of our methods:
\subsubsection{IM- with Race groups and Age-associated risk-weights (IM-RA)} We perform equal treatment to achieve representative allocation of vaccines amongst racial groups, but use the influence function $\sigma_A$ to apply a heavy penalty for infecting older communities.

\subsubsection{IM - with median Income and Age-associated risk-weights (IM-IA)} Similarly, we perform equal treatment of vaccines amongst the four income groups and replace the influence function with $\sigma_A$.
In our experiments, we later find high numbers of submodularity violations when using the $\sigma_A$ influence function. For all experiments using the ``-A'' suffix, we, therefore, omit the recalculation checks in Lines 10 to 15 of Algorithm \ref{alg:im}. 

\section{Experiments}
\noindent \textbf{Dataset}
We conduct experiments on three mobility networks of MSAs in the US, constructed from individual mobile tracking data from SafeGraph, from the Dewey platform \cite{Safegraph}. We use the implementation proposed by \cite{Chang2020} to construct these networks, and also use their Covid-19 model as our influence function. 
A mobility network is constructed as a temporal bipartite graph $G^t=(V,E^t)$, whose nodes $V$ are either CBGs, which are communities of between 600 and 3,000 US residents, or POIs.
A directed weighted edge $e^t(v_1,v_2) \in E^t$ represents the number of residents from a CBG who are visiting a POI at hour $t$. The graph varies over time, such that no nodes are added or removed, but the edge weights vary hourly. We use five weeks of visits from CBGs to POIs beginning on the 2nd of March 2020. For our implementation, only the first two weeks are used to select the most influential communities, and vaccination is implemented after the two-week mark. We used data from the beginning of March in order to exploit the full mobility of individuals before lockdown. Otherwise, it would not have been possible to measure the effect of the proposed vaccination strategies separately from the effect of lockdown.

Throughout the experiments, we set the vaccination budget to 5\% of the population size of the whole network. We experiment with three MSAs - Philadelphia, New York and Chicago - each of which encompasses the main city as well as the wider metropolitan area. These particular MSAs were selected based on their high racial diversity, and high discrepancy in infections between racial groups as reported in the previous work. The outputs are mobility networks modelling metropolitan areas of populations between 6 and 10 million residents. We provide details of further pre-processing steps and network statistics in the full paper.\\

\noindent \textbf{Baselines}
Each vaccination strategy is run over 30 random seeds, and we report the average results in comparison to three baseline approaches. We design our own baselines here, since the closest approach to ours which uses IM for vaccination uses a generic propagation function which is not specific to a particular disease, and is valid only on rooted trees and not general graphs \cite{Minutoli2020preempt}.

\begin{itemize}
 \item \textit{No vaccination:} We model free-spreading Covid-19 during the total five-week period, without any vaccination strategy.
 \item  \textit{Random vaccination (RAND):} We implement a random selection of CBGs for vaccination within the budget $B$. We collect results over three random seeds and report the average.
 \item \textit{Current strategy proxy (CS):} Here we replicate the current strategy which prioritizes older communities. We select the oldest communities for vaccination, by the median age of the CBG, up to the vaccine budget $B$.
\end{itemize}

\noindent \textbf{Fairness Evaluation Metrics}
In addition to evaluating the performance of the vaccination strategies, we propose two methods of evaluating fairness for social groups in this context. In both cases, we measure the discrete KL-divergence between two distributions; we compare a distribution from the outcome of our experiments $p(j)$ to an ideal ``fair'' distribution $q(j)$. We draw on fairness notions from the IM literature - equal treatment and equal outcomes. For both measures, the fair distribution $q(j)$ corresponds to the fractions of each social group in the network, $q(j) = N_j/N$. 

\begin{itemize}
    \item \textbf{Equal treatment}
For equal treatment, we aim to obtain a fair representation of each social group $j$ within the CBGs selected for vaccination. As such, the output distribution $p(j)$ is the proportion of social group $j$ amongst vaccinated CBGs, $p(j) = (N_j/N)_V$.

    \item \textbf{Equal outcome}
To obtain equal outcomes, the goal is to ensure that no individual is more at risk of infection than the fraction of that social group in the network dictates. In this case, we set $p(j)$ to be the proportion of infections received by social group $j$, as a result of our vaccination strategy. This can be written as $p(j) = (N_j/N)_{EIR}$.
\end{itemize}

\section{Results and Discussion}
\begin{figure}
    \centering
    \includegraphics[width=8.4cm]{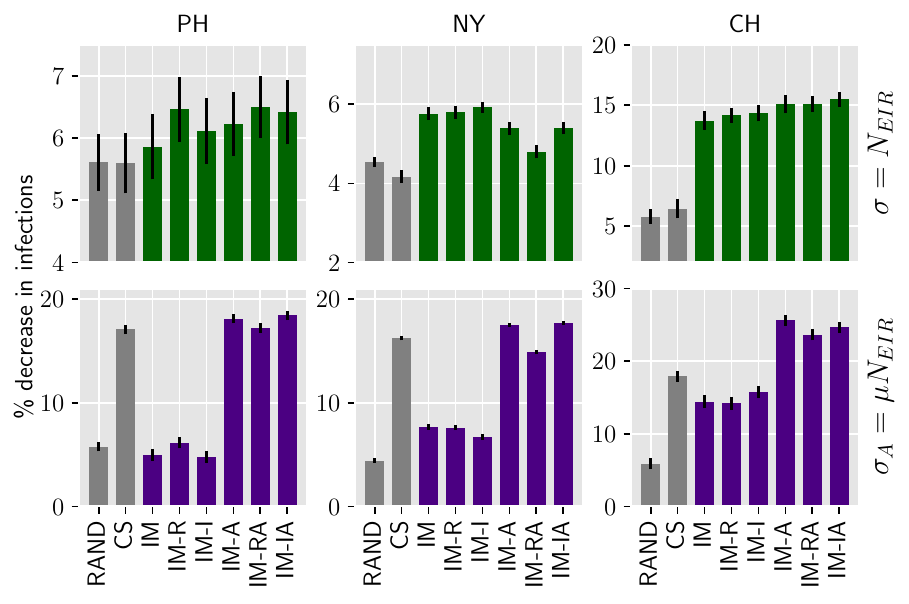}
    \caption{Performance measured by percentage decrease in infections (top), and percentage decrease in risk-weighted infections, i.e. with a weighted penalty of infecting older communities (bottom), compared to not vaccinating. Higher is better for both metrics.}
    \label{fig:performance}
\end{figure}

\subsubsection{Reducing overall infections}
The performance of each vaccination strategy in reducing the number of infections is shown in Figure \ref{fig:performance} (green). For every MSA, all vaccination strategies using IM outperform both RAND and CS. Though the infections decrease by only a few percent, given the network size, these percentage point differences are significant. For example, a 5\% decrease in infections for Philadelphia corresponds to around 18,000 fewer people infected (including estimates of unreported infections). We observe that the variations of IM experiments which include fairness (the last five bars) do not experience a significant decrease in performance even when optimizing for both performance \textit{and} fairness. This illustrates that fairness considerations in vaccination distribution do not have to come at the cost of increasing infection counts.

\subsubsection{Infections in high-risk groups}
Figure \ref{fig:performance} (purple) reports the percentage decrease in infections weighted by age-associated risk, as described in Equation \ref{eq:risk}. The experiments optimizing for age are the baseline CS and our contributions IM-A, IM-RA, and IM-IA, which therefore perform best for this metric. However, we see that all of our proposed solutions which optimize for age outperform the current strategy, even when they are also optimizing for another sensitive attribute at the same time. The results testify to several alternative solutions that better protect older communities as well as ensure fairness for other sensitive attributes like race and income.

\begin{figure}
    \centering
    \includegraphics[width=8.4cm]{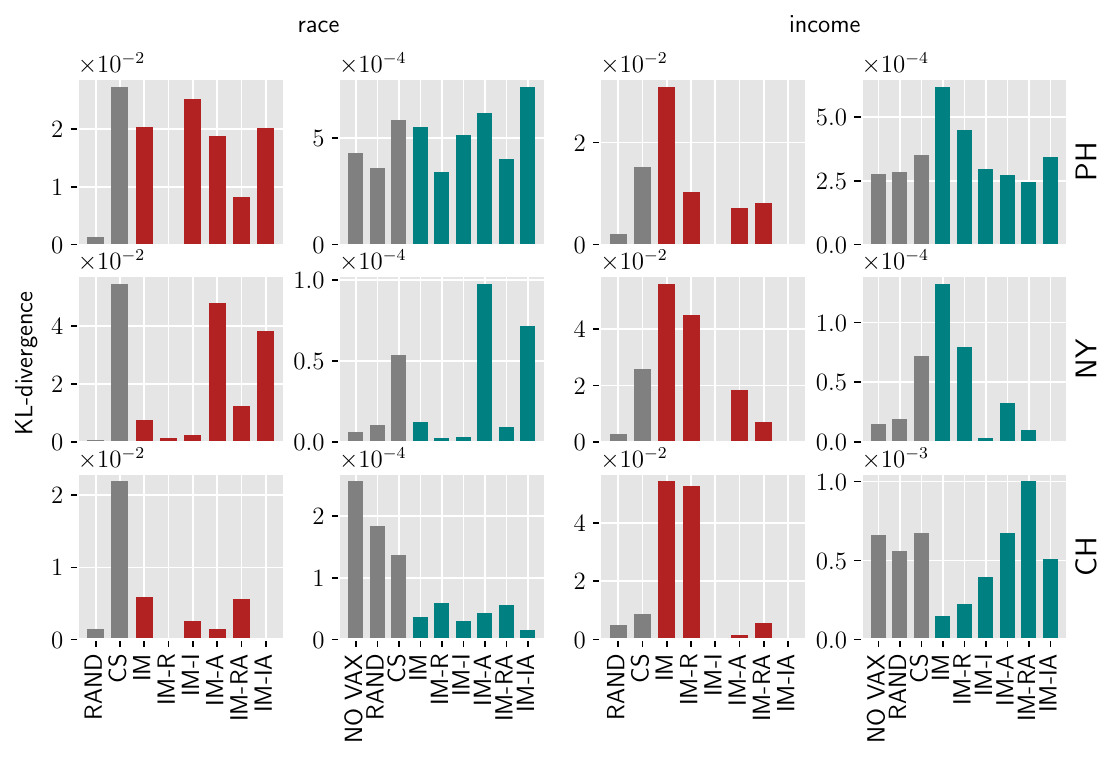}
    \caption{The KL-divergence scores measure fair treatment (red) and fair outcomes (blue) with respect to racial groups (left) and income groups (right). Lower $D_{KL}$ corresponds to better fairness for both metrics.}
    \label{fig:all-fairness}
\end{figure}

\subsubsection{Comparing fairness notions}
In Figure \ref{fig:all-fairness} left, we present results for equal treatment (red) and equal outcome (blue) for racial groups, and in Figure \ref{fig:all-fairness} right we present the same metrics for income groups. Though the $D_{KL}$ values are small, these still correspond to significant differences in these large networks. For example, the current strategy model (CS) has a $D_{KL}$ of around $5.8 \times 10^{-4}$ for Philadelphia, which corresponds to the Black population suffering around 6,000 more infections than if they were distributed according to their proportion in the network. Critically, we see that IM-R methods successfully achieve equal treatment for racial groups across all MSAs, as does IM-RA, with the exception of Chicago. The effect is stronger for income, where IM-I and IM-IA experiments achieve near-perfect ($D_{KL}$ of zero) distribution of vaccines to income groups, for all MSAs. Though equal treatment is important to ensure, we are more interested in an equal outcome, as infections are the more serious consequence. We see that the experiments which perform best for equal outcomes amongst races are also IM-R and IM-RA. This demonstrates that achieving equal treatment can be very effective in delivering fair outcomes for those same demographic groups. Additionally, for some MSAs, constraining on fairness by \textit{income} (IM-I) leads to fair outcomes for \textit{race}, and vice versa. This implies that the objectives of achieving fairness amongst races and fairness amongst social status are similar.

\subsubsection{Optimizing for multiple sensitive attributes}
Since the results of every vaccination experiment for each MSA can differ, it is possible that there is no one one-size-fits-all best vaccination strategy for every urban area. In particular, we can  identify how accounting for higher-risk individuals (experiments with the ``-A'' suffix) can work favourably for achieving demographic fairness in infections for some but not all MSAs (see Chicago, Figure \ref{fig:all-fairness} blue, left and New York, Figure \ref{fig:all-fairness} blue, right). Despite this, we can identify at least one strategy per metropolitan area which achieves high performance in reducing infections overall, as well as a competitive result for the infection outcomes of \textit{all three} sensitive attributes of age, race, and social status by income: IM-RA for Philadelphia and New York, and IM-IA for Chicago.

\section{Conclusion}
For policy-makers, choosing a vaccination strategy amongst those presented here is non-trivial. There is no one-size-fits-all solution for every urban area. However, we demonstrate that, for all networks we tested here, one of our proposed methods can successfully ensure demographic fairness for all three sensitive attributes. We, therefore, argue that community-level influence maximization should be incorporated into whichever ethical stance is taken, and we present the methodology to do so.

Our approach can be extended to accommodate multiple rounds of vaccine allocation, as commonly observed in real-world scenarios. It would be necessary to capture the mobility shifts at different stages of lockdown, and how this affects demographic groups differently. There are many factors which could be incorporated to make the simulation more realistic, such as vaccine hesitancy. For later stages of vaccine roll-out, our approach could be combined with data on hesitancy rates to prioritize neighborhoods with low up-take, while maintaining demographic fairness.

\newpage
\section*{Acknowledgements}
Funding support for project activities has been partially provided by Canada CIFAR AI Chair, Facebook award, MEI award, and NSERC Discovery Grants program. We also express our gratitude to Compute Canada for their support in providing facilities for our evaluations.

\bibliography{references}
\newpage
\appendix
\section{Constructing Mobility Networks} \label{app:networks}
We use the Covid-19 simulation and mobility network proposed by Chang et al. \cite{Chang2020}. To construct the networks, they used the Safegraph social-distancing data of daily estimates of the fraction of CBG's residents who are out visiting other CBGs. Safegraph no longer provides these values, so we estimated it ourselves using Safegraph's Neighborhood Patterns of visitors \textit{arriving at} CBGs, and aggregating over these. We make our code available upon publication of this work to demonstrate our approach. Additionally, due to sparsity in the weekly patterns data used to construct the mobility networks, we aggregate over two previous months of the monthly patterns data. We use fewer months of aggregate visits than the previous approach, which means our networks' statistics differ from theirs. This also leads to Covid-19 model parameters which differ from that of the previous work (see Table \ref{tab:parameters}).

\begin{table}[h]
    \centering
    \begin{tabular}{cccl}
        \hline
        & $\beta_{\mathrm{home}}$ & $\psi$ & $p_0$ \\
        \hline
       Philadelphia & 0.02 & 300 & 0.001 \\
       New York & 0.02 & 100 & 0.005 \\
       Chicago & 0.02 & 500 & 0.0005 \\
       \hline
    \end{tabular}
    \caption{Final model parameters obtained by tuning infection model to real Covid-19 case counts for each MSA.}
    \label{tab:parameters}
\end{table}

Table \ref{tab:graphstats} summarizes the network statistics of the three MSAs in terms of population, CBGs, and POIs. 
\begin{table}[h]
\centering
  \begin{tabular}{cccl}
    \hline
    MSA & Population & CBGs & POIs \\
    \hline
   Philadelphia & 9,247,281 & 5,603 & 11,479 \\
   New York & 9,990,617 & 6,522 & 20,606 \\
   Chicago & 6,074,364 & 3,452 & 8,281 \\
 \hline
\end{tabular}
\caption{Final statistics of mobility networks.}
\label{tab:graphstats}
\end{table}

\section{Covid-19 Model}
Below we describe the basic set-up of the Covid-19 model. More details can be found in the supplementary information of the previous work by Chang et al. \cite{Chang2020}. The model maintains four vectors of size $K$ - $S$, $E$, $I$ and $R$ - corresponding to the fraction of susceptible, exposed, infected and recovered/removed individuals per CBG. The rates of transitions between these states at a time step $t$ are determined by the number of new exposures $N^{(t)}_{S_{c_i}\rightarrow E_{c_i}}$, the number of exposures transitioning to infections $N^{(t)}_{E_{c_i}\rightarrow I_{c_i}}$, and the number of infections transitioning to recovered/removed, $N^{(t)}_{I_{c_i}\rightarrow R_{c_i}}$. 
The number of new exposures depends on two factors; visits to public places (from the visit matrix $w$) containing other infectious individuals, and interactions from the home CBG with other infectious individuals. Transitions are sampled from the following distributions:
\begin{equation}
    N^{(t)}_{S_{c_i}\rightarrow E_{c_i}} \sim \mathrm{Pois} \frac{S^{(t)}_{c_i}}{N_{c_i}} \sum^n_{j=1} \lambda^{(t)}_{p_j}w^{(t)}_{ij} + \mathrm{Binom}(S_{c_i}^{(t)}, \lambda_{c_i}^{(t)})
\end{equation}
\begin{equation}
    N^{(t)}_{E_{c_i}\rightarrow I_{c_i}} \sim \mathrm{Binom}(E^{(t)}_{c_i}, 1/\delta_E)
\end{equation}
\begin{equation}
    N^{(t)}_{I_{c_i}\rightarrow R_{c_i}} \sim \mathrm{Binom}(I^{(t)}_{c_i}, 1/\delta_I)
\end{equation}
Here, $\lambda_{p_j}$ refers to the infection rate at POI $p_j$, and $\lambda_{c_i}$ is the infection rate at CBG $c_i$. $\delta_E$ and $\delta_I$ refer to the mean exposure period and the mean infectious period respectively. Using the number of transitions, the number of $S$, $E$, $I$ and $R$ for a CBG $c_i$ at time $t>0$ can be expressed as follows:
\begin{equation}
    S_{c_i}^{(t)} = S_{c_i}^{(t-1)} - N^{(t)}_{S_{c_i}\rightarrow E_{c_i}}
\end{equation}
\begin{equation}
    E_{c_i}^{(t)} = E_{c_i}^{(t-1)} - N^{(t)}_{E_{c_i}\rightarrow I_{c_i}} + N^{(t)}_{S_{c_i}\rightarrow E_{c_i}}
\end{equation}
\begin{equation}
    I_{c_i}^{(t)} = I_{c_i}^{(t-1)} + N^{(t)}_{E_{c_i}\rightarrow I_{c_i}} - N^{(t)}_{I_{c_i}\rightarrow R_{c_i}}
\end{equation}
\begin{equation}
    R_{c_i}^{(t)} = R_{c_i}^{(t-1)} + N^{(t)}_{I_{c_i}\rightarrow R_{c_i}}
\end{equation}
At $t=0$: 
\begin{equation}
      E_{c_i}^{(0)} = \left\{\begin{matrix}
      N_{c_i} p_0 && \text{if $c_i$ is infected}
\\ 
0 && \text{otherwise.}
\end{matrix}\right.
\end{equation}
\begin{equation}
    S_{c_i}^{(0)} = N_{c_i} - E_{c_i}^{(0)}
\end{equation}
\begin{equation}
    I_{c_i}^{(0)} = 0
\end{equation}
\begin{equation}
    R_{c_i}^{(0)} = 0
\end{equation}
where $p_0$ is the probability that an individual in the metro area is exposed at the first time step. This probability is one of the three parameters set when calibrating the disease propagation to real case counts, see Table \ref{tab:parameters}. 
\section{Age group risk factors}
Table \ref{tab:risks} shows the different risk factors of infection, hospitalisation, and death of different age brackets compared to 18-29 years olds. The trend shows that while younger people are more prone to getting infected, they are less likely to face grave or fatal consequences. For instance, a person aged 85 is  350 times more likely to die from covid compared to a person in their 20s. It is important to consider such risks when using developing a vaccination strategy.

\begin{table}[h]
\centering
  \begin{tabular}{ccccccc}
    \hline
     & 30-39 & 40-49 & 50-64 & 65-74 & 75-84 & 85+ \\
    \hline
    C & 1.0x & 0.9x & 0.8x & 0.6x & 0.6x & 0.7x \\
   H & 1.5x & 1.9x & 3.1x & 4.8x & 8.6x & 15x \\
   D & 3.5x & 10x & 25x & 60x & 140x & 350x \\
 \hline
\end{tabular}
\caption{Age group risk factors of Cases, Hospitalization and Death compared to 18-29 year olds, from CDC \cite{cdc-risks}.}
\label{tab:risks}
\end{table}

\section{Infrastructure}
To first construct the three mobility networks, we use 2CPUs and 64Gb of memory on Linux OS. The vaccination experiments can then each be run on 1CPU, 2GPU and 8Gb memory. We point the reader to the README file included in the code repository for a list of libraries and packages required to run on conda environments.
The code can be found here: github.com/nicolaneo/fair\_vaccination\_with\_im.

\end{document}